Title: **"Carbon Chain Depletion of 2I/Borisov"**
Author List:
**Theodore Kareta[1], Jennifer Andrews[2], John W. Noonan[1], Walter M. Harris[1], Nathan Smith[2], Patrick O'Brien[1], Benjamin N. L. Sharkey[1], Vishnu Reddy[1], Alessandra Springmann[1], Cassandra Lejoly[1], Kathryn Volk[1], Albert Conrad[3], Christian Veillet[3]**
1: Lunar and Planetary Laboratory
2: Steward Observatory
3: Large Binocular Telescope Observatory

Abstract: The composition of comets in the Solar System come in multiple groups thought to encode information about their formation in different regions of the outer protosolar disk. The recent discovery of the second interstellar object, 2I/Borisov, allows for spectroscopic investigations into its gas content and a preliminary classification of it within the Solar System comet taxonomies to test the applicability of planetesimal formation models to other stellar systems. We present spectroscopic and imaging observations from 2019 September 20th through October 26th from the Bok, MMT, and LBT telescopes. We identify CN in the comet's spectrum and set precise upper limits on the abundance of $C_2$ on all dates. We use a Haser model to convert our integrated fluxes to production rates and find Q(CN) = 5.0 +/- 2.0 * 10^24 mol/s on September 20[th] and Q(CN) = (1.1 – 1.9) * 10^24 mols/s on later dates, both consistent with contemporaneous observations. We set our lowest upper limit on a $C_2$ production rate, $Q(C_2)$ < 1.6 10^23 mols/s on October 10th. The measured ratio upper limit for that date $Q(C_2)$/Q(CN) < 0.095 indicates that 2I/Borisov is strongly in the (carbon chain) 'depleted' taxonomic group. The only comparable Solar System comets have detected ratios near this limit, making 2I/Borisov statistically likely to be more depleted than any known comet.
Most 'depleted' comets are Jupiter Family Comets, perhaps indicating a similarity in formation conditions between the most depleted of the JFCs and 2I/Borisov. More work is needed to understand the applicability of our knowledge of Solar System comet taxonomies onto interstellar objects and we discuss future work that could help to clarify the usefulness of the approach.

## Introduction

The modern composition and structure of Solar System comets retains information about conditions of the early protosolar disk from which they formed (Duncan, Quinn, and Tremaine, 1987; Bar-Nun and Kleinfield, 1989; Mumma, Weissman, and Stern 1993; A'Hearn et al., 2012). Telescopic surveys of comet molecular abundances (A'Hearn et al., 1995; Fink 2009; Cochran et al., 2012) have revealed there to be multiple taxonomic classes of Solar System comets whose differences are likely tied to differences in disk chemistry. Dynamical work also suggests multiple formation regions inside the disk (Biver and Bockelée-Morvan, 2015; Dones et al., 2015; Bockelée-Morvan and Biver, 2017; Meech, 2017) related to the modern dynamical classes of comets (Jupiter Family, Halley-type, Long Period, etc.). Connections between dynamical class and composition ("normal", carbon-chain depleted, carbon-chain and $NH_2$ depleted, etc.) have also been identified (A'Hearn et al., 1995; Fink 2009; Cochran et al., 2012) using spectroscopy of cometary comae. Spectroscopic observations of an active cometary coma, whether the object is from our Solar System or interstellar, thus investigates the link between disk properties and produced planetesimals, and provides constraints on models of planetesimal and planet formation.

The discovery of the second-ever interstellar object comet 2I/Borisov on 2019 August 30, (MPEC 2019-R106) presents a unique opportunity to contrast planetesimal formation in our Solar System with formation processes in other stellar systems—in a way not possible with the first interstellar object found, 1I/'Oumuamua. 'Oumuamua, is red with a pronounced elongation (Meech et al., 2017; Knight et al., 2017; Fitzsimmons et al., 2018; Ye et al., 2018; Trilling et al., 2018) with an assumed low, but undetermined, visible (< 0.15) and infrared (< 0.2) albedo (Fitzsimmons et al., 2018; Trilling et al., 2018) in a non-principal axis ("tumbling") rotational state (Drahus et al., 2018; Fraser et al., 2018, Belton et al., 2018), and strong evidence for non-gravitational acceleration (Micheli et al., 2018). No gas emissions were detected from 'Oumuamua: chemical analyses were limited to observing its surface reflectivity (Fitzsimmons et al., 2018) and are consistent with an outer Solar System body with appreciable organic content. All of these properties contrast with 2I/Borisov; the object's substantial outgassing provides more data about its formation, thermal history, and chemistry.

We summarize the currently known properties of 2I/Borisov. Its eccentricity and inclination are 3.31º and 44.1º, respectively (JPL Orbit Solution #11, queried for 2019 October 1.0). Due to its low solar elongation angle it is visible only in the early morning, presenting a brief observing window. Guzik et al., (2019) obtained g'- and r'-band images with Gemini North, finding 2I/Borisov's color to be indistinguishable from Solar System comets. de Léon et al., (2019) reported a red slope between 0.55–0.90 µm similar to D-type asteroids and cometary nuclei. Fitzsimmons et al., (2019) detected CN emission from 2I/Borisov (Q ~ 4 x $10^{24}$ molecules/s) at a level appearing typical for comets at similar heliocentric distances. Fitzsimmons et al., (2019) also constrained on $C_2$ production (Q ≤ 4 x $10^{24}$ molecules/s) and estimated water production by proxy (Q ~ 1.7 x $10^{27}$ molecules/s). **McKay et al. (2019) derived from the 6300 Angstrom [OI] line a water production rate of ~6.3 *10^26 mols/s, which could be high or typical compared to solar system comets depending on size estimated (Jewitt and Lu, 2019).**

Carbon "typical" comets have production rate ratios of $C_2$/CN of 0.66–3.0 (A'Hearn et al., 1995; Cochran et al., 2012), while comets with less $C_2$ are called 'depleted'. Fitzsimmons et al., (2019) cannot yet discriminate 2I/Borisov between 'typical' and 'depleted', though the detection of CN does effectively rule out an uncommon composition like that of Comet Yanaka (1988r; Fink, 1992) or 96P/Machholz (Schleicher, 2008). **Opitom et al. (2019) found no evidence of $C_2$ in their spectroscopic observations, suggesting moderate-to-strong ($C_2$/CN < 0.3) carbon-chain depletion.** Spectroscopic observations can place chemical constraints on 2I/Borisov, allowing insight into another Solar System's disk chemistry and formation history by examination of the interstellar comet's $C_2$ and CN abundance in ways that 1I/'Oumumua could not.

## Observations & Data Reduction

We observed interstellar comet 2I/Borisov on 2019 September 20 with the Boller & Chivens spectrograph at the 2.3-m Bok telescope (Kitt Peak National Observatory, Arizona), on 2019 October 1 **and 9** with the Blue Channel spectrograph at the 6.5-m MMT telescope (Mount Hopkins, Arizona) (Hastie and Williams, 2010), **and on 2019 October 10 with MODS on the Large Binocular Telescope (LBT, Pogge et al., 2006)** to search for and measure the abundance of molecule species as well as to measure the reflective properties of dust in the object's coma. **We also imaged Borisov in the Hale-Bopp CN (Farnham, Schleicher, and A'Hearn, 2000) and r' filters on 2019 October 26 with the 90Prime imager (Williams et al., 2004) on the Bok telescope.** Observational details, **including heliocentric and geocentric distances, spectral resolution, and exposure times among other information,** are listed in **Table 1**. For context, our Bok observations took place approximately 7 hours after Fitzsimmons et al. (2019).

**Table 1 - Observing Circumstances for 2I/Borisov**

| Telescope | Date | $R_H$ (au) | Δ (au) | Time (UTC) | Airmass | $T_{Exp}$ (s) | Angstrom/pix | Conds. |
|---|---|---|---|---|---|---|---|---|
| Bok (2.3-m) | Sep. 20 | 2.67 | 3.25 | 12:07-12:17 | 1.89 | 600 | 3.60 | Photometric |
|  | Oct. 26 | 2.23 | 2.52 | 12:19-12:29 | 1.46 | 1200 | HB CN Filter | Photometric |
| MMT (6.5-m) | Oct. 1 | 2.50 | 3.00 | 11:47-11:53 | 1.87 - 1.93 | 2 x 300 | 1.95 | Cirrus? |
|  | Oct. 9 | 2.41 | 2.84 | 11:46-12:06 | 1.75-1.82 | 4 x 300 | 1.95 | Cirrus? |
| LBT (2x8.4m) | Oct. 10 | 2.39 | 2.82 | 12:00-12:16 | 1.57-1.64 | 2x | 0.5 | Photometric |

**Airmasses listed are for the start of each exposure.**

These spectra **and images** were reduced using standard techniques, including bias subtraction, flat fielding, cosmic ray rejection, local sky subtraction and extraction of one-dimensional spectra. The spectra had wavelength solutions derived using via comparison with Helium-Neon-Argon lamps, had a standard extinction correction applied, and were flux-calibrated by comparison with nearby flux standard stars observed just beforehand. **For the spectroscopic observations extracted apertures were chosen to maximize comet signal for each instrument and slit. For the data from the 20[th], this was an 18.33" (spatial) by 1.5" (slit width) box, the 1[st] had 15.575" by 1.0", the 9[th] had 26.0" by 1.5", and the 10[th] had 18" by 2.4". The imaging data was extracted using circular aperture photometry with a radius of 25.0" and largely followed the reduction procedure outlined in Farnham, Schleicher, and A'Hearn (2000) using the dust reflectance slope measured from our Oct. 10[th] spectrum. The standard star used (HD 81809) is noted in that work to be less than** ideal for blue/UV observations, so we have overestimated errors when possible. We note that for the Sep. 20 and Oct. 1 datasets, insufficient data was collected to allow a median combination of spectra, making the two datasets "look" even noisier than their later counterparts even though cosmic rays were cleaned from them.

## Analysis, Modeling, and Results

In order to search for molecular emission features, it is necessary to produce a flux-calibrated solar spectrum fit to subtract from the flux-calibrated data. This process creates a dust-reflectance spectrum for the comet, which can be used to look for a spectral slope. The solar spectrum was taken from **Chance and Kurucz (2010) and binned to the resolution of each dataset**. The original flux calibrated spectra, best-fit reflectance, and dust-subtracted comet spectra are shown in **Figure 1**. The noisiness of the Bok data prevents a high-quality solar reflectance subtraction, while the later data become very linear after subtraction indicating a relatively good fit of the solar spectrum to the continuum points.

**We consider the spectral slope between the blue, green, and red continuum regions defined for the Hale-Bopp filter set (Farnham, Schleicher, and A'Hearn, 2000) to compare reflectance slopes with other workers. If we normalize our highest-quality data from October 10$^{th}$ at the green continuum point, then we find the slope from blue-to-green to be S' = 22 +/- 4 %/micron, and the green-to-red slope to be S' = 11 +/- 3 %/micron. Fitzsimmons et al. (2019) suggested that the dust might be 'redder' at shorter wavelengths, and our best data supports that conclusion.**

**Figure 1 - Spectral Observations of 2I/Borisov**

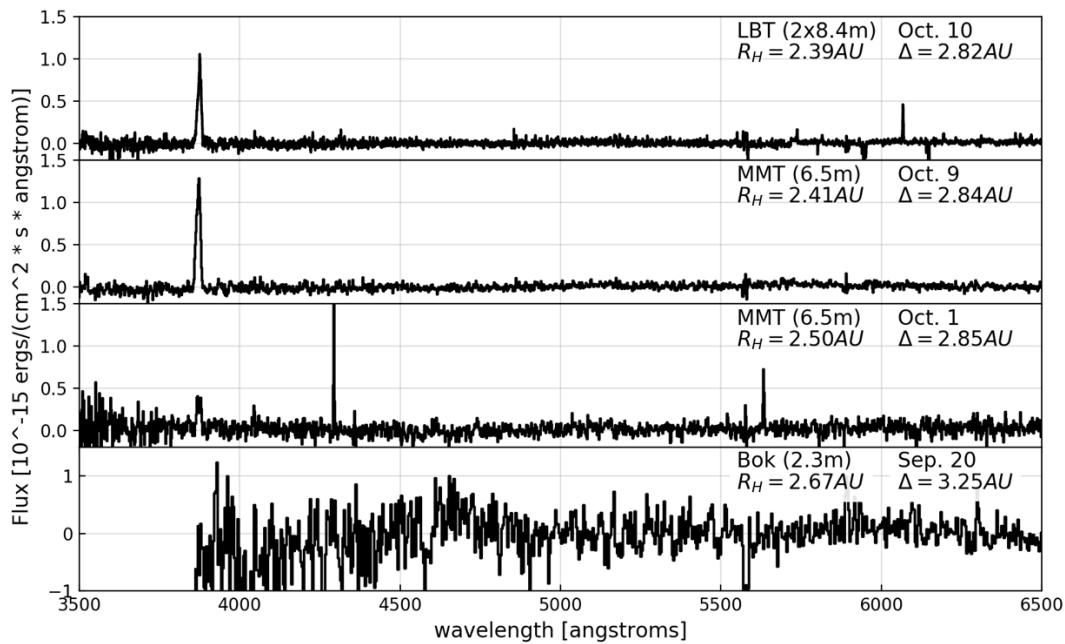

**Figure 1 Caption**: The flux-calibrated and dust-subtracted spectra of **all four nights of spectroscopy**. The heliocentric and geocentric distances are listed underneath the telescope names and dates on the figure. The same background and background subtraction process was applied to **all** datasets. The 3880.0 Å CN feature can be seen clearly in the MMT **and LBT** data.

After the reflected sunlight from dust was subtracted, we searched for emission from molecules in the coma of 2I/Borisov (**Table 2**). For each possible feature we subtracted a linear estimate of the background as determined by a fit to data on either side of the feature and then integrated using a trapezoidal method all flux above it. That retrieved flux is compared against an upper limit found through the method described in Cochran et al., (2012), whereby a fraction of the local spectral noise is multiplied over the width of the bandpass that the feature was integrated over to determine blindly if the feature is 'real'. All retrieved fluxes and upper limits are in **Table 2**.

The CN (0-0) band near 3880 Å is a relatively narrow two-peaked feature which we integrated from 3861 Å to 3884 Å based on the extent of the feature as calculated in Schleicher (2010). **We detect CN strongly on Oct. 1, 9, and 10, but the detection on Sep. 20 is marginal, as background subtraction was challenging so close to the edge of the detector. We thus list it in Table 2 as an both an upper limit and a detection.**

The $C_2$ Swan band ($\Delta \nu = 0$), with a bandhead at 5167.0 Å, is a much broader feature stretching over 100 Å in wavelength. The overall shape of the band has a sharp peak near the bandhead and a secondary smoother peak near 5100 Å. We integrate from 50**67** Å to the bandhead at 5167 Å based on previous observations of Solar System comets **and to match the procedures of Fitzsimmons et al. (2019) and Opitom et al. (2019)**. There is no $C_2$ seen **above the noise in any of the data.** Detailed sections of the **October 1st, 9th, and 10th** spectra of 2I/Borisov show the CN **detections** and $C_2$ non-detections on **those dates** (**Figure 2**).

To convert the integrated fluxes to production rates, we used a standard Haser model (Haser 1957) with scale lengths and lifetimes from A'Hearn et al., (1995) and outflow velocity from Cochran and Schleicher (1993). The g-factor (formally the fluorescence efficiency) for $C_2$ was taken from A'Hearn et al., (1995) and the g-factor for CN was taken from Schleicher (2010) for the appropriate heliocentric **distance and** velocity on each date. The lifetimes increased proportional to heliocentric distance squared and the g-factors were scaled down by the same factor. We have verified the output of our Haser model implementation by comparison of its output with a recent paper on the topic, Hyland et al. (2019). The input fluxes and output production rates **for all nights** are presented in **Table 2**. The production rate ratio **upper limit**, $Q(C_2)/Q(CN)$, is measured as **<0.095** which indicates that 2I/Borisov is clearly in the 'depleted' group of the A'Hearn et al. (1995) taxonomy. A comparison of our inferred $Q(C_2)/Q(CN)$ ratio with those of the A'Hearn et al. (1995) sample as submitted to the Planetary Data System in 2003 (Osip et al., 2003) **as well as that of Opitom et al. (2019)** is presented in **Figure 3**.

**Table 2 - Fluxes and Production Rates of Cometary Molecules**

| Telescope | Flux (CN) | Flux (C2) | Q(CN) | Q(C2) |
|---|---|---|---|---|
| Bok (Sep. 20) | <= (1.6 +/- 0.6) * 10^-14 | < 1.8 *10^-14 | <= (5.0 +/- 2.0) * 10^24 | < 8.0 * 10^24 |
| MMT (Oct. 1) | (2.9 +/- 0.4) * 10^-15 | < 4.4 * 10^-15 | (1.1 +/- 0.2) * 10^24 | < 2.5 * 10^24 |
| MMT (Oct. 9) | (9.1 +/- 0.5) * 10^-15 | < 1.8 * 10^-15 | (1.59 +/- 0.09) * 10^24 | < 4.4 * 10^23 |
| LBT (Oct. 10) | (1.19 +/- 0.03) * 10^-14 | < 8.0 * 10^-16 | (1.69 +/- 0.04) * 10^24 | < 1.62 * 10^23 |
| Bok (Oct. 26) | (3.8 +/- 0.7) * 10^-13 | | (1.9 +/- 0.3) * 10^24 | |

Fluxes are in ergs/cm²/s, and production rates are in molecules/s.

**Figure 2 - CN and C$_2$ spectral profiles at the MMT and LBT.**

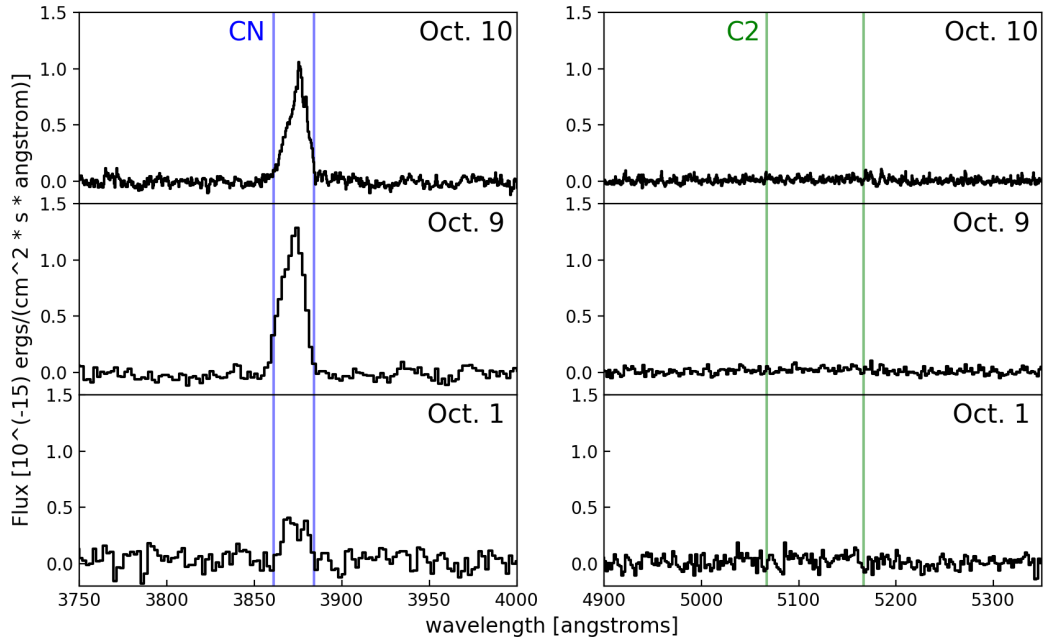

**Figure 2 Caption:** The CN (left) and C$_2$ (right) spectral profiles from the spectra obtained between October 1st and October 10th. Dotted horizontal lines are added to each plot to indicate the extraction spectral range, **which we chose based on previous work in the field.**

**Figure 3 - Taxonomical Classification of 2I/Borisov**

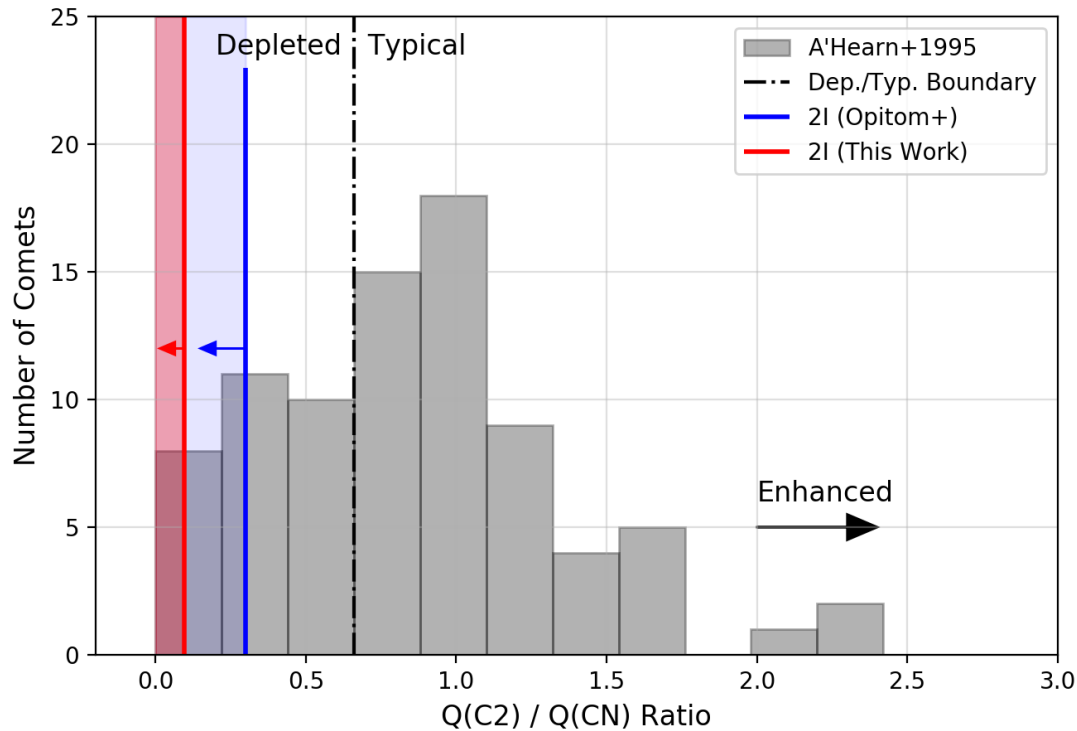

**Figure 3 Caption**: The ratio of production rates between $C_2$ and CN determined for 2I/Borisov **(red line, red shaded area, red arrow)** and the comets in the A'Hearn et al. (1995) survey which had their ratios determined (Osip et al., 2003). **The upper limit from Opitom et al., (2019) is presented similarly in blue.** A black dash-dotted line and text is present to differentiate between 'Depleted' comets (ratio < 0.66) and 'Typical' comets (ratio > 0.66) in the A'Hearn et al. (1995) taxonomy. 'Enhanced' comets fall on the right and outside of the figure.

## Discussion

**2I/Borisov as a Depleted Comet**

Our **further** confirmation of CN emission, **continuing non-**detection of $C_2$ emission, and **lowest so far** calculation of the $Q(C_2)/Q(CN)$ ratio provide a better framework than previously available to analyze 2I/Borisov in the context of Solar System comets. Our reported $Q(C_2)/Q(CN)$ **upper limit (<0.095) is very low for solar system comets, with only three in the A'Hearn et al. (1995) sample being comparably depleted (43P/Wolf-Harrington: ~0.063, 98P/Takamizawa: ~0.081, and 87P/Bus: ~0.087). Only ~34% of the total comets with measured $Q(C_2)/Q(CN)$ ratios in that sample are depleted. We note that these detections are rather noisy and comparatively old.** $Q(C_2)/Q(CN)$ has also been observed to vary in 43P (Fink 2009) from very to mildly depleted. 19P/Borelly also has a variable ratio (Fink 2009) that on one occasion was as low as ~0.06. All of these lowest-ever detections are comparable to our upper limit, making 2I/Borisov's statistically likely to be more depleted than any solar system comet.**

In our Solar System, Jupiter Family Comets are the most likely to be depleted (~37% vs. ~18% for long period comets, Cochran et al., 2012; ~50% vs. **~6%** in A'Hearn et al. 1995) depending on definition of depleted and choice of dynamical classifications. Within this context, we might expect 2I/Borisov to have formed in a location more similar to many of the **extremely** **depleted** Jupiter Family Comets, derived from the trans-Neptunian region, than the Long Period Comets, which likely formed near Uranus and Neptune. Inferences like this rest upon the assumption that the existing classification schemes are describing some underlying set of processes in the protosolar disk that created the modern properties and compositions of comets. More work is needed to understand if 2I/Borisov being carbon-depleted has the same implications as it does in Solar System comets. A reliable detection of the molecule $C_2$ in the spectrum of 2I/Borisov would help address this. $C_2$ and $C_3$ are usually seen to be depleted, typical, or enriched in a similar manner (A'Hearn et al., 1995), suggesting strongly that their abundance is controlled by the same or similar processes. If $C_3$ is detected and is found to be correlated in production rate with $C_2$, this would likely argue further that the taxonomies reflect some information about the chemical conditions in the disks where these objects formed. Further spectroscopic monitoring of 2I/Borisov is needed for another reason; while production rate ratios (e.g. $Q(C_2)/Q(CN)$) are not observed to vary much with time or distance in Solar System comets, we do not know if this is the case for 2I/Borisov as of yet. If these diagnostic ratios are found to vary in 2I/Borisov, then our understanding of what our measurement of its moderate depletion means would have to be revised greatly, if not thrown out altogether.

**Comparison with Other Measurements**

Our September 20th measurements took place approximately 7 hours after those of Fitzsimmons et al. (2019) and our Q(CN) values rates agree within 1-sigma (Q(CN) = 3.7 ± 0.4 * 10^24, Fitzsimmons et al., 2019, Q(CN) = 5.0 ± 2.0 * 10^24, this work.) adding credence to our marginal detection. **Our later observations result in lower CN production rates ((1.2 - 1.9) *10^24), which are wholly consistent with the results recently reported by Opitom et al., 2019. Considering the differences in instruments, reduction strategies, and modeling approaches, we view this as good agreement with contemporary work that bolsters the validity of our approaches. Understanding how production rates vary with heliocentric distance is a critical goal for the upcoming observational campaign, as any deviations from normal cometary behavior might indicate some different mechanisms that regulate the release of gases from the interior of the object. While Opitom et al. (2019) argue based on their data that the trend in Q(CN) is essentially flat, we note that while our data supports such an idea, it could also support a weak increase over the same period.** We argue that the current data supports the idea that 2I/Borisov is acting much like a normal Solar System comet in this regard.

**Both Fitzsimmons et al. (2019) and Opitom et al. (2019) placed upper limits on the $Q(C_2)/Q(CN)$ ratio (~1.0 and ~0.3, respectively). Our upper limit is significantly below both, but our derived Q(CN) production rates are statistically consistent. We view this as strong evidence that our derived upper limit is accurate.**

**Our dust reflectance slope measurements (S' = 22%/micron and 11%/micron below and above the green continuum) support the idea of Fitzsimmons et al. (2019) that the dust might be more 'red' at shorter wavelengths. However, we note that small differences between measured reflectance slopes can be caused by a number of things, from different calibration methods to unknown instrumental artifacts, so more measurements are needed. A recent report (Yang et al., 2019, CBET 4672) reports a somewhat shallower slope (~5%), for instance.** This is a typical Solar System value (c.f. Guzik et al. 2019) and indicates dust with relatively similar reflective properties as that of Solar System comets. **The dust properties and general behavior of 2I/Borisov seem similar to many Solar System comets, but its gas abundances might be very different.**

# Conclusions

**We report on spectroscopic and imaging observations of 2I/Borisov between 2019 September 20$^{th}$ and 2019 October 26$^{th}$ using the Bok, MMT, and LBT telescopes to characterize and measure the gas production rates and dust reflectance properties to compare with Solar System comets. We find the dust reflectance to be red (S'=22 (11) %/micron) and likely shallower at larger wavelengths. We identify CN emission on all dates and convert the integrated fluxes to production rates using a standard Haser model with a rectangular (for spectra) or circular (for imaging) aperture. From October 1$^{st}$ to 26$^{th}$, the production rate ranged from Q(CN) = 1.1 – 1.9 * 10^24 mols/s after our initial marginal detection on September 20$^{th}$ (Q(CN) = 5.0 +/- 2.0 * 10^24 mol/s). These values and dust reflectance properties are all consistent with contemporaneous observations by Fitzsimmons et al. (2019) and Opitom et al. (2019). We do not detect emission from $C_2$ on any date and set upper limits in the standard method of Cochran et al. (2012). For our most sensitive observations from the LBT on October 10$^{th}$, we find Q($C_2$) < 1.6 * 10^23. The ratio of $C_2$ to CN production rates was found on that date to be Q($C_2$)/Q(CN) < 0.095, which is extremely depleted. This ratio is used as a diagnostic measure of composition (e.g. 'typical' or 'depleted', see A'Hearn et al., 1995, Cochran et al., 2012, Fink 2009), and only three Solar System comets are more depleted, which suggests that Borisov likely has a very rare composition by Solar System standards.** However, more work is needed before we can truly apply a Solar System classification onto 2I/Borisov and know that the statement has a physical meaning. **More work is very much needed to further constrain this ratio and identify if 2I/Borisov has any long-chain hydrocarbons at all.** The recent discovery of the interstellar comet 2I/Borisov presents an unparalleled opportunity to compare its composition to Solar System comets and test the applicability of planetesimal formation models to other stellar systems.

## Acknowledgements

We would like to congratulate Gennadiy Borisov on this great discovery and thank the IAU for letting the object retain his name as is typical for comets. We thank Kendall Sullivan among others for encouragement to pursue this project. **We thank an anonymous reviewer for comments that greatly improved the paper as well as Alan Fitzsimmons for very helpful correspondence on the topic.** JEA and NS receives support from NSF grant AST-1515559. **KV acknowledges funding from NSF (grant AST-1824869).** Some observations in this work were collected at Kitt Peak National Observatory; we are honored to be permitted to conduct astronomical research on Iolkam Du'ag (Kitt Peak), a mountain with particular significance to the Tohono O'odham Nation. Observations reported here were obtained at the MMT Observatory, a joint facility of the University of Arizona and the Smithsonian Institution. We appreciate the expertise of the Kitt Peak National Observatory, Steward Observatory, MMT Observatory, and LBT Observatory staff as well as Hannes Gröller who helped ensure the success of these observations.

This paper made use of the modsIDL spectral data reduction reduction pipeline developed in part with funds provided by NSF Grant AST-1108693 and a generous gift from OSU Astronomy alumnus David G. Price through the Price Fellowship in Astronomical Instrumentation. modsCCDRed was developed for the MODS1 and MODS2 instruments at the Large Binocular Telescope Observatory, which were built with major support provided by grants from the U.S. National Science Foundation's Division of Astronomical Sciences Advanced Technologies and Instrumentation (AST-9987045), the NSF/NOAO TSIP Program, and matching funds provided by the Ohio State University Office of Research and the Ohio Board of Regents. Additional support for modsCCDRed was provided by NSF Grant AST-1108693.

## Facilities
MMT(BCS), Bok (B&C, 90Prime), LBT (MODS)

## Bibliography

- A'Hearn, M. F., Millis, R. C., Schleicher, D. O., Osip, D. J., & Birch, P. V. 1995, Icarus, 118, 223
- Biver, N. and Bockelée-Morvan, D., 2015. Chemical diversity in the comet population. *Proceedings of the International Astronomical Union*, *11*(A29A), pp.228-232.
- Bar-Nun, Akiva, and Idit Kleinfeld. "On the temperature and gas composition in the region of comet formation." *Icarus* 80.2 (1989): 243-253.
- Bockelée-Morvan, D., Crovisier, J., Mumma, M.J. and Weaver, H.A., 2004. The composition of cometary volatiles. *Comets II*, *1*, pp.391-423.
- Bockelée-Morvan, D. and Biver, N., 2017. The composition of cometary ices. *Philosophical Transactions of the Royal Society A: Mathematical, Physical and Engineering Sciences*, *375*(2097), p.20160252.
- Belton, M. J. S., Hainaut, O. R., Meech, K. J., et al. 2018, ApJL, 856, L21
- Cochran, A. L., Barker, E. S., & Gray, C. L. 2012, Icarus, 218, 144
- Dones, Luke, et al. "Origin and evolution of the cometary reservoirs." *Space Science Reviews* 197.1-4 (2015): 191-269.
- Drahus, M., Guzik, P., Waniak, W., Handzlik, B., Kurowski, S., & Xu, S. (2018). Tumbling motion of 1I/'Oumuamua and its implications for the body's distant past. *Nature Astronomy*, *2*(5), 407.
- Duncan, Martin, Thomas Quinn, and Scott Tremaine. "The formation and extent of the solar system comet cloud." *The Astronomical Journal* 94 (1987): 1330-1338.
- Fraser, W.C., Pravec, P., Fitzsimmons, A., Lacerda, P., Bannister, M.T., Snodgrass, C. and Smolić, I., 2018. The tumbling rotational state of 1I/'Oumuamua. *Nature Astronomy*, *2*(5), p.383.
- Fink, U., 1992. Comet Yanaka (1988r): A new class of carbon-poor comet. *Science*, *257*(5078), pp.1926-1929.
- Fink, U., 2009. A taxonomic survey of comet composition 1985–2004 using CCD spectroscopy. *Icarus*, *201*(1), pp.311-334.
- Fitzsimmons, A., Snodgrass, C., Rozitis, B., et al. 2018, Nature Astronomy, 2, 133
- Hastie, M., and George G. Williams. "Instrumentation suite at the MMT Observatory." *Ground-based and Airborne Instrumentation for Astronomy III*. Vol. 7735. International Society for Optics and Photonics, 2010.
- Irvine, WILLIAM M., et al. "Comets: A link between interstellar and nebular chemistry." *Protostars and planets IV* 1159 (2000).
- Knight, Matthew M., et al. "On the rotation period and shape of the hyperbolic asteroid 1I/'Oumuamua (2017 U1) from its lightcurve." *The Astrophysical Journal Letters* 851.2 (2017): L31.
- León, Julia De, et al. "Interstellar Visitors: A Physical Characterization of Comet C/2019 Q4 (Borisov) with OSIRIS at the 10.4 m GTC." *Research Notes of the AAS*, vol. 3, no. 9, 2019, p. 131., doi:10.3847/2515-5172/ab449c.
- Meech, K. J., Weryk, R., Micheli, M., et al. 2017, Nature, 552, 378
- Micheli, M., Farnocchia, D., Meech, K.J., Buie, M.W., Hainaut, O.R., Prialnik, D., Schörghofer, N., Weaver, H.A., Chodas, P.W., Kleyna, J.T. and Weryk, R., 2018. Non-gravitational acceleration in the trajectory of 1I/2017 U1 ('Oumuamua). *Nature*, *559*(7713), p.223.
- Mitchell, G. F., J. L. Ginsburg, and P. J. Kuntz. "A steady-state calculation of molecule abundances in interstellar clouds." *The Astrophysical Journal Supplement Series* 38 (1978): 39-68.
- Mumma, Michael J., Paul R. Weissman, and S. A. Stern. "Comets and the origin of the solar system-Reading the Rosetta Stone." *Protostars and planets III*. 1993
- Schleicher, D.G., 2008. The extremely anomalous molecular abundances of comet 96P/Machholz 1 from narrowband photometry. *The Astronomical Journal*, *136*(5), p.2204.
- Stecher, Theodore P., and David A. Williams. "Interstellar molecule formation." (1966).

- Trilling, D. E., Mommert, M., Hora, J. L., et al. 2018, The Astronomical Journal, 156, 261
- Ye, Q.-Z., Zhang, Q., Kelley, M. S. P., & Brown, P. G. 2017, The Astrophysical Journal, 851, L5
- **Pogge, R. W., Atwood, B., Belville, S. R., Brewer, D. F., Byard, P. L., DePoy, D. L., Derwent, M. A.,Eastwood, J., Gonzalez, R., Krygier, A., Marshall, J. R., Martini, P., Mason, J. A., O'Brien, T. P., Osmer,P. S., Pappalardo, D. P., Steinbrecher, D. P., Teiga, E. J., and Weinberg, D. H., "The multi-object double spectrographs for the Large Binocular Telescope," in [Society of Photo-Optical Instrumentation Engineers(SPIE) Conference Series],Proc. SPIE6269, 62690I (June 2006).**
- **Williams, G. G., Olszewski, E., Lesser, M. P., & Burge, J. H. 2004, Ground-based Instrumentation for Astronomy, 5492, 787**
- **Chance, K., & Kurucz, R. L. 2010, Journal of Quantitative Spectroscopy and Radiative Transfer, 111, 1289**
- Opitom, C., Fitzsimmons, A., Jehin, E., et al. 2019, arXiv e-prints, arXiv:1910.09078. accepted for publication in A&A Letters.
- 

Guzik, P., Drahus, M., Rusek, K., et al. 2019, arXiv:190905851 [astro-ph], http://arxiv.org/abs/1909.05851
Fitzsimmons, A., Hainaut, O., Meech, K., et al. 2019, arXiv:190912144 [astro-ph], http://arxiv.org/abs/1909.12144
Yang, B. et al., CBET 4672
Jewitt, D., & Luu, J. 2019, arXiv e-prints, arXiv:1910.02547